# Nonlinearities and Noise-Signal Relations in Electronic Heat Transport via Molecules

Arthur Luniewski, Rita Aghjayan, Kamil Walczak

*Department of Chemistry and Physical Sciences, Pace University*
*1 Pace Plaza, New York, NY 10038, USA*

(Dated: August 1, 2016)

We examine the electronic heat transport phenomena in nanoscale junctions composed of organic molecules coupled to two metallic reservoirs of different temperatures. The electronic heat flux and its dynamical noise properties are calculated within the scattering (Landauer) formalism with the transmission probability determined by using non-equilibrium Green's functions (NEGF technique). The method based on Taylor series expansion is used to determine nonlinear corrections to the electronic heat flux and its noise power spectral density with up to the second order terms with respect to the temperature difference. Our results show only limited applicability of ballistic Fourier's law and fluctuation-dissipation theorem to heat transport in molecular systems. We derived and tested numerically several signal-signal, noise-signal and noise-noise relations applicable to nanoscale heat flow carried by electrons at strongly non-equilibrium conditions (similar formulas are expected for phonons and photons). Importantly, the special treatment proposed by us may be extended to higher order terms in order to address a variety of problems related to nonlinear thermal and electro-thermal effects which may occur at nanoscale.

## I. INTRODUCTION

The Fourier's physics for heat flow is based on two fundamental postulates. The *first Fourier's law* interrelates heat flux $J$ with temperature difference $\Delta T$ in the ballistic heat transport regime [1], or its vector $\vec{J}$ with temperature gradient $\vec{\nabla} T$ in the diffusive heat transport regime [2]

$$J = K \Delta T, \quad \textbf{(ballisticity)} \tag{1}$$

$$\vec{J} = -\kappa A \vec{\nabla} T. \quad \textbf{(diffusivity)} \tag{2}$$

Here $K$ is a size-dependent (extensive) quantity known as thermal conductance, $\kappa$ is material-specific (intensive) quantity known as thermal conductivity, while $A$ is the cross-sectional area associated with heat conduction process. Those transport laws are strictly related to the linear response regime and may be violated in the case of high-intensity heat fluxes, in the presence of time-dependent thermal effects, and because of confinement as well as quantization effects occurring at nanoscale. The *second Fourier's law* is a scaling rule for the length-dependent thermal conductance

$$K = \text{const}(L), \quad \textcolor{blue}{\textbf{(ballisticity)}} \tag{3}$$

$$K = \kappa \frac{A}{L}. \quad \textcolor{red}{\textbf{(diffusivity)}} \tag{4}$$

Here $L$ is the length of the conduction channel. It should be noted that the second Fourier's law may be broken in biological and mesoscopic systems, where certain scattering events, collective phenomena, as well as correlation effects become important. The experimental verifications of Fourier's law is usually based on Eqs.(3) and (4), where $K \sim L^{\beta-1}$ or $\kappa \sim L^{\beta}$ with $\beta = 0$ for diffusive heat transfer or $\beta = 1$ for ballistic heat flow [3-6].

In nanoscale systems, transport of energy carried by electrons turned out to be quantized [7-10] and may be a strongly nonlinear function of temperature difference. From quantum mechanical point of view, we have learnt how to treat nonlinear transport properties of molecular complexes and the associated noises in terms of Landauer conduction channels within the following set of equations [11,12]

$$J = \frac{1}{2\pi\hbar} \sum_n \int_{-\infty}^{+\infty} (\varepsilon - \mu) P_n(\varepsilon) [f_L - f_R] d\varepsilon, \tag{5}$$

$$S_{TH} = \frac{1}{\pi\hbar} \sum_n \int_{-\infty}^{+\infty} (\varepsilon - \mu)^2 P_n(\varepsilon) [f_L(1 - f_L) + f_R(1 - f_R)] d\varepsilon, \tag{6}$$

$$S_{SN} = \frac{1}{\pi\hbar} \sum_n \int_{-\infty}^{+\infty} (\varepsilon - \mu)^2 P_n(\varepsilon) [1 - P_n(\varepsilon)] (f_L - f_R)^2 d\varepsilon. \tag{7}$$

The summation in Eqs.(5)-(7) is over each conduction channel which contributes independently to electronic heat transport. Here $S_{TH}$ denotes equilibrium thermal noise, $S_{SN}$ stands for non-equilibrium shot noise of quantum nature, $P_n(\varepsilon)$ is probability that an electron of energy $\varepsilon$ will pass through nanoscale system, $f_L$ and $f_R$ are Fermi-Dirac distribution functions which capture statistics of electrons in the left and right reservoirs, respectively. However, the proper definition of Landauer conduction channels remains an open question in nanoscale physics [13].

In this paper, we use nonlinear heat transport theory applicable to molecular junctions composed of a single molecule coupled to two metallic reservoirs (thermal baths). Specifically, we focus our attention on three major aspects of nanoscale heat conduction. (1) Landauer conduction channels which we define as discrete energy levels of a molecule characterized by delocalized orbitals and significant overlap with reservoir continuum of states (relatively strong reservoir-molecule coupling). (2) Molecular noises which may mask original signals (heat fluxes), but may also be analyzed, providing some useful information about individual heat carriers in the system under investigation, e.g. their statistical independence or correlation effects. Here we propose a new type of noise-signal relations which will be tested numerically. (3) Nonlinear corrections to heat fluxes and the associated noises in the ballistic heat transport regime. It should be noted that Eqs.(5)-(7) become questionable when we are dealing with channel mixture effects in which electron passing through one energy level is scattered into a different one. It is particularly true in molecular systems, where the phase coherence of the electronic wave functions is maintained during heat propagation. This channel mixture and other interference effects call for transport formalism with globally defined transmission function (not as the sum of individual Landauer conduction channels), where all the interference and overlapping phenomena are included, while heat carriers from different channels do interfere and must remain indistinguishable.

## II. ELECTRONIC HEAT FLUX

We analyze heat transport phenomena within a conceptually simple and transparent Landauer formalism which relates transport properties of nanoscale system to its scattering properties, such as globally defined quantum-mechanical transmission probability function. Knowing the energy-dependent transmission function for electrons $P(\varepsilon)$, we can use the Landauer-type formula to calculate the so-called electronic heat flux [14,15]

$$J = \int_{-\infty}^{+\infty} P(\varepsilon) F(\varepsilon, T) d\varepsilon, \tag{8}$$

$$F(\varepsilon, T) \equiv \frac{(\varepsilon - \mu)}{2\pi\hbar}[f_L - f_R]. \tag{9}$$

Here $f_L = f_L(\varepsilon, T + \Delta T)$ and $f_R = f_R(\varepsilon, T)$ are Fermi-Dirac occupation factors for the left (L) and right (R) thermal baths (heat reservoirs), T is absolute temperature, $\Delta T$ is temperature difference between two reservoirs. For practical reasons, we expand Eq.(9) into Taylor series with respect to temperature difference $\Delta T$ (it is usually a small parameter) to obtain expression for the heat flux in the form

$$J = \sum_{m=1}^{\infty} K^{(m)} (\Delta T)^m, \tag{10}$$

where the particular transport coefficients are defined via the following integral

$$K^{(m)} = \int_{-\infty}^{+\infty} P(\varepsilon) F_m(\varepsilon, T) d\varepsilon. \tag{11}$$

The integral kernels in Eq.(11) are temperature-dependent functions which for the first two terms take the form

$$F_1(y) = \frac{k_B y^2}{8\pi\hbar} \cosh^{-2}\left(\frac{y}{2}\right), \tag{12}$$

$$F_2(y) = \frac{k_B y^2}{8\pi\hbar T} \cosh^{-2}\left(\frac{y}{2}\right)\left[\frac{y}{2}\tanh\left(\frac{y}{2}\right) - 1\right], \tag{13}$$

where variable y is defined as

$$y(\varepsilon, T) = \frac{\varepsilon - \mu}{k_B T}. \tag{14}$$

Obviously, we could expand Eq.(9) to higher-order terms with respect to temperature difference $\Delta T$ (see the Appendix A), but in this paper we restrict ourselves only to the most essential quadratic corrections. In Eq.(10), the coefficient $K^{(1)}$ plays the role of a linear thermal conductance, while $K^{(2)}$ defines the first nonlinear coefficient in the modified Fourier's law.

To give an example, let us consider a special case of ballistic heat transport regime, where the heat is transferred by a single energy level without any scattering involved into the conduction process of electrons. In this case $P(\varepsilon) = 1$, and after performing integration in Eq.(11), we obtain

$$J = \frac{\pi k_B^2 T}{6\hbar} \Delta T \left[ 1 + \frac{1}{2} \left\{ \frac{\Delta T}{T} \right\} \right] \equiv K^{(1)} \Delta T + K^{(2)} (\Delta T)^2. \tag{15}$$

According to Eq.(15), in the ballistic heat transport regime, when the scattering processes are negligible, each energy level contributes a universal quantum $\pi k_B^2 / 6\hbar$ to the reduced thermal conductance $K^{(1)}/T$. At strongly non-equilibrium conditions, when the temperature difference $\Delta T$ is significant in comparison to other energy parameters, there is additional universal correction $\pi k_B^2 \Delta T / 12\hbar$ to the electronic thermal conductance $K^{(1)}$ (this is actually the only non-zero correction in the ballistic heat transport regime). Although the amount of energy transferred by individual electrons is not quantized, thermal conductance turned out to be a quantized quantity. Quantization of thermal conductance may also be understood as a consequence of Heisenberg uncertainty principle as combined with equipartition theorem from thermodynamics (see the Appendix B). In Fig.1, we plot nonlinear (quadratic) ballistic Fourier's law which is a linear function of temperature $T$ and quandratic function of temperature difference $\Delta T$. Interestingly, the quanta of thermal conductance and their nonlinear corrections associated with phonons [16-18] and photons [19,20] are expressed by exactly the same formulas as Eq.(15).

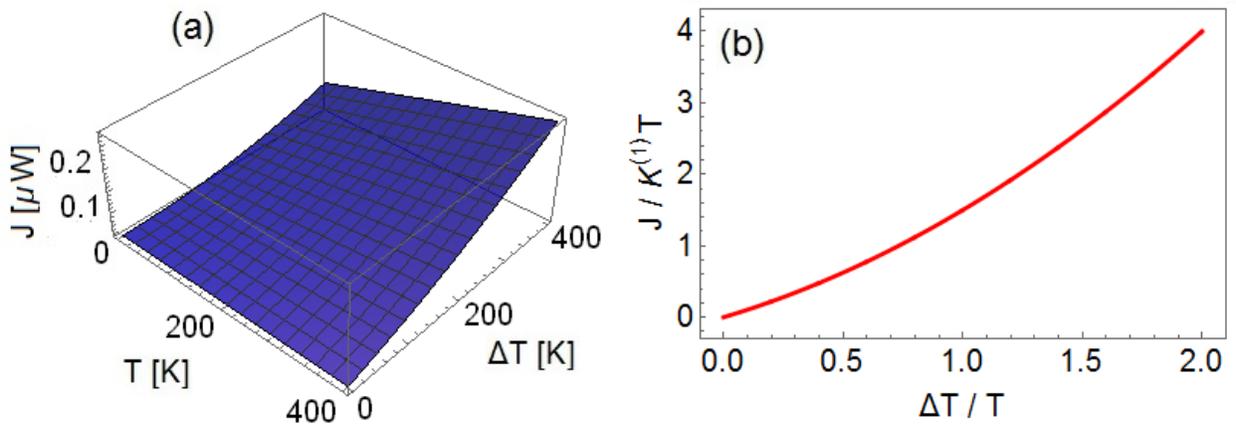

**Figure 1:** (a) 3D visualization of nonlinear (quadratic) ballistic Fourier's law; (b) 2D visualization of nonlinear (quadratic) ballistic Fourier's law.

## III. NOISES AND NOISE-SIGNAL RELATIONS

In general, the electronic heat flux is fluctuating in time. The noise associated with nanoscale transport of conducting electrons is usually described by the symmetrized autocorrelation function of the form [21,22]

$$S(t) = \frac{1}{2}\langle \delta J(t)\delta J(0) + \delta J(0)\delta J(t) \rangle. \tag{16}$$

Here the angular brackets indicate an ensemble average, while the variance (uncertainty) of the energy flux in the presence of a heat flow carried by individual electrons is defined as

$$\delta J(t) = J(t) - \langle J(t) \rangle. \tag{17}$$

In the asymptotic limit of infinite time ($t \to \infty$), the frequency-dependent power spectrum density related to heat flow carried by electrons is simply a Fourier transform of the autocorrelation function from Eq.(16), namely

$$S(\omega) = \frac{1}{2\pi} \int_{-\infty}^{+\infty} S(t)\exp(-i\omega t) dt. \tag{18}$$

Using the Landauer scattering formalism in the zero-frequency limit ($\omega \to 0$), the noise power of electronic heat flux fluctuations may be conveniently decomposed into two separate parts: the equilibrium (Johnson-Nyquist) thermal noise and non-equilibrium (Schottky) shot noise

$$S = S_{TH} + S_{SN}, \tag{19}$$

$$S_{TH} = \int_{-\infty}^{+\infty} P(\varepsilon)\Phi(\varepsilon, T) d\varepsilon, \tag{20}$$

$$S_{SN} = \int_{-\infty}^{+\infty} P(\varepsilon)[1 - P(\varepsilon)]\Psi(\varepsilon, T) d\varepsilon, \tag{21}$$

where the integral nuclei in Eqs.(20) and (21) are expressed via the following relations

$$\Phi(\varepsilon, T) = \frac{(\varepsilon - \mu)^2}{\pi\hbar}\left[f_L(1-f_L) + f_R(1-f_R)\right], \tag{22}$$

$$\Psi(\varepsilon, T) = \frac{(\varepsilon - \mu)^2}{\pi\hbar}\left[f_L - f_R\right]^2. \tag{23}$$

Since integral in Eq.(21) is strictly semipositive, the non-equilibrium term can only enhance the noise above the level established by the equilibrium thermal fluctuations. The expansion of Eq.(22) into the Taylor series with respect to temperature difference $\Delta T$ allows us to write down the following expression for the thermal noise

$$S_{TH} = \sum_{m=0}^{\infty} S_{TH}^{(m)} (\Delta T)^m. \tag{24}$$

The particular noise coefficients in Eq.(24) are defined via the following integral

$$S_{TH}^{(m)} = \int_{-\infty}^{+\infty} P(\varepsilon) \Phi_m(\varepsilon, T) d\varepsilon. \tag{25}$$

The integral kernels in Eq.(25) are temperature-dependent functions which for the first three terms take the form

$$\Phi_0(x) = \frac{(k_B T y)^2}{\pi \hbar} [\cosh(y) + 1]^{-1}, \tag{26}$$

$$\Phi_1(x) = \frac{k_B^2 T y^3}{4\pi \hbar} \cosh^{-2}\left(\frac{y}{2}\right) \tanh\left(\frac{y}{2}\right), \tag{27}$$

$$\Phi_2(x) = \frac{k_B^2 y^3}{8\pi \hbar} \cosh^{-4}\left(\frac{y}{2}\right) \left[\frac{y}{2}[\cosh(y) - 2] - \sinh(y)\right], \tag{28}$$

where the variable y is defined once again via Eq.(14). Since Eq.(24) is expressed via infinite series with the zeroth-order term, the equilibrium thermal noise is unavoidable even in the absence of temperature difference $\Delta T$ (or temperature gradient $\vec{\nabla} T$), what is fully consistent with our intuition.

As an example, let us consider a special case of ballistic heat transport regime, where the heat is carried by electrons passing through the single energy level without any scattering processes involved into the conduction process. In this case $P(\varepsilon) = 1$, and after performing integration in Eq.(25), we obtain

$$S_{TH} = \frac{2\pi k_B^3 T^3}{3\hbar} \left[1 + \frac{3}{2}\left\{\frac{\Delta T}{T}\right\} + \frac{3}{2}\left\{\frac{\Delta T}{T}\right\}^2 + ...\right] \equiv S_{TH}^{(0)} + S_{TH}^{(1)} \Delta T + S_{TH}^{(2)} (\Delta T)^2 + ... \tag{29}$$

At thermal equilibrium, both reservoirs are kept at the same temperature ($\Delta T = 0$) and the average heat flux is zero $\langle J(t) \rangle = 0$, but the equilibrium thermal noise $S_{TH}^{(0)}$ is directly proportional to the linear thermal conductance $K^{(1)}$ and may still be significant

$$S_{TH}^{(0)} = 4 k_B T^2 K^{(1)}. \tag{30}$$

Eq.(30) is a simple manifestation of fluctuation-dissipation theorem (FDT) [23-26], where the noise component is related to the signal (transport coefficient). This relation is usually regarded as a universal equation, which is assumed to be independent of materials, chemical nature of interatomic connections, and the specific models used in simulations of transport characteristics associated with thermal devices. However, in our analysis related to ballistic heat transport regime, we can find two similar expressions for higher-order thermal noises

$$S_{TH}^{(1)} = 6k_B T K^{(1)},\qquad(31)$$

$$S_{TH}^{(2)} = 6k_B K^{(1)}.\qquad(32)$$

Furthermore, we can also propose three analogous expressions for the considered thermal noise components by involving nonlinear correction to thermal conductance term $K^{(2)}$, namely

$$S_{TH}^{(0)} = 8k_B T^3 K^{(2)},\qquad(33)$$

$$S_{TH}^{(1)} = 12k_B T^2 K^{(2)},\qquad(34)$$

$$S_{TH}^{(2)} = 12k_B T K^{(2)}.\qquad(35)$$

The obvious question is related to the universal character of all the noise-signal relations given by Eqs.(30)-(35). Further, calculating specific ratios of thermal noises related to different order terms, we can prove the following expressions

$$S_{TH}^{(0)} = \frac{2}{3} T S_{TH}^{(1)} = \frac{2}{3} T^2 S_{TH}^{(2)}.\qquad(36)$$

The importance of Eqs.(36) stem from the fact that one can deduce the higher-order terms of equilibrium thermal noise on the basis of its zeroth-order component (which is usually easily measurable quantity at nearly equilibrium conditions). In Section V, we provide numerical evidence that the fluctuation-dissipation theorem from Eq.(30) is universally valid for molecular systems and arbitrary temperatures of heat reservoirs, but is limited to small temperature difference between two thermal baths (as a consequence of linear response theory). Further, we show that all noise-signal relations from Eqs.(31)-(35) and noise-noise relations from Eqs.(36) are applicable to molecular systems in the transport regime when temperature difference between heat reservoirs is significant, while the average temperature of the system under consideration is relatively small (in comparison to room temperature).

In the next step, let us analyze non-equilibrium shot noise. The expansion of Eq.(23) into the Taylor series with respect to temperature difference $\Delta T$ allows us to write down the following expression

$$S_{SN} = \sum_{m=2}^{\infty} S_{SN}^{(m)} (\Delta T)^m.\qquad(37)$$

Here the particular noise coefficients are defined via the following integral transform

$$S_{SN}^{(m)} = \int_{-\infty}^{+\infty} P(\varepsilon)[1 - P(\varepsilon)] \Psi_m(\varepsilon, T) d\varepsilon.\qquad(38)$$

The integral kernels in Eq.(38) are temperature-dependent functions. In the absence of temperature difference ($\Delta T = 0$), the non-equilibrium shot noise is identically equal to zero $S_{SN}^{(0)} = 0$. Further, we can show that the linear term with respect to the finite temperature difference ($\Delta T \neq 0$) also disappears $S_{SN}^{(1)} = 0$, and the first non-zero term is defined as a quadratic correction to the shot noise with the integral kernel of the form

$$\Psi_2(x) = \frac{k_B^2 y^4}{16\pi\hbar} \cosh^{-4}\left(\frac{y}{2}\right). \tag{39}$$

It should be noted that the non-equilibrium shot noise disappears for ideal ballistic conduction, for which $P(\varepsilon) = 1$, and reaches maximum value for $P(\varepsilon) = 1/2$. We can evaluate the maximum value of the non-equilibrium shot noise related to a single energy level by putting $P(\varepsilon) = 1/2$ into Eq.(38) and after performing integration, we obtain

$$S_{SN,max}^{(2)} \cong \frac{k_B^3 T}{4\pi\hbar}. \tag{40}$$

It this special case, we can come up with additional two equations relating the maximum value of non-equilibrium shot noise with transport coefficients

$$S_{SN,max}^{(2)} = \frac{k_B}{2} K^{(1)}, \tag{41}$$

$$S_{SN,max}^{(2)} = k_B T K^{(2)}. \tag{42}$$

Although Eqs.(41) and (42) look very similar to fluctuation-dissipation relations, their limited applicability is rather obvious, since they are valid only for specific energies $\varepsilon$ for which electron has 50% chance to be transmitted $P(\varepsilon) = 1/2$ and 50% chance for being reflected $[1 - P(\varepsilon)] = 1/2$. Further, by comparing Eq.(41) to Eq.(42) or equivalently Eq.(30) to Eq.(33), or Eq.(31) to (34) or even Eq.(32) to Eq.(35), we obtain formula which connects linear thermal conductance with its quadratic correction

$$K^{(1)} = 2T K^{(2)}. \tag{43}$$

In Section V, we provide numerical evidence that the applicability of signal-signal relation from Eq.(43) is limited to significantly large temperature difference between heat reservoirs, while the average temperature of the system under consideration is still relatively small.

## IV. ELECTRONIC TRANSMISSION FUNCTION

In order to use the perturbative transport theory developed in Section III, and to test numerically all the signal-signal, signal-noise and noise-noise relations introduced there, we need to define transmission function for the system composed of molecule coupled to two heat reservoirs which are kept at different temperatures. The usual choice is Caroli formula for the transmission probability expressed in terms of the effective propagators (Green's functions) [27]

$$T(\varepsilon) = \text{Trace}[\Gamma_L G \Gamma_R G^+]. \tag{44}$$

Here $\Gamma_L$ and $\Gamma_R$ are broadening functions of the contacts, while G is an effective propagator (Green's function) which captures the information about the electronic structure of the molecule as specifically modified by its connection to heat reservoirs. This propagator is a solution of the so-called Dyson equation

$$[\varepsilon I - H - \Sigma_L - \Sigma_R] \, G(\varepsilon) = I. \tag{45}$$

Here $\varepsilon$ is the energy of heat carrier, I is the unity matrix, H is Hamiltonian of molecular system, while $\Sigma_L$ and $\Sigma_R$ are self-energy terms related to broadening functions via the following expressions

$$\Gamma_\alpha(\varepsilon) = i[\Sigma_\alpha - \Sigma_\alpha^+] = \tau_\alpha A(\varepsilon) \tau_\alpha^+, \tag{46}$$

Here index $\alpha = L, R$ stands for the left and right reservoirs, $A(\varepsilon)$ is the energy-dependent surface spectral function, while $\tau_\alpha$ is the matrix describing molecular connection to heat reservoirs, respectively. The self-energy terms can be expressed as follows

$$\Sigma_\alpha(\varepsilon) = \tau_\alpha Q(\varepsilon) \tau_\alpha^+ = \Lambda_\alpha(\varepsilon) - \frac{i}{2} \Gamma_\alpha(\varepsilon). \tag{47}$$

Here the surface propagator $Q(\varepsilon)$ include the information about the electronic structure of thermal baths, while the real part of self-energy term $\Lambda_\alpha(\varepsilon)$ and the broadening function $\Gamma_\alpha(\varepsilon)$ are interrelated via the so-called Hilbert transform [28]. It should be noted that the real part of self-energy term describes the shift of discrete energy levels of the molecule itself, while its imaginary part is responsible for contact-induced broadening of those energy levels.

## V. RESULTS AND DISCUSSION

So far, the presented formalism was very general and many different levels of description may be used to study heat transfer at nanoscale. However, in this paper, we limit ourselves to a very simple models applicable to molecular junction composed of benzene molecule coupled to two metallic (golden) reservoirs via thiol groups, as schematically shown in Fig.2. The organic molecule itself is treated within the tight-binding approaximation, where only delocalized pi-orbitals with hopping parameters $t = -2.7$ eV are included into our simplified Hamiltonian (all onsite energies has been made equal to zero) [29]. In the case of single-atom connection between molecule and individual heat reservoir, the coupling matrix $\tau_\alpha$ is reduced to only one energy parameter $\tau$. Both heat reservoirs are modeled via semi-elliptical density of states [30]

$$Q(\varepsilon) = \frac{1}{\gamma}\left[\frac{\varepsilon}{2\gamma} - i\sqrt{1 - \left(\frac{\varepsilon}{2\gamma}\right)^2}\right], \tag{48}$$

with $\gamma = 10$ eV to mimic wide-band metal and electrochemical potential or Fermi energy level $\mu = 0$ (this is the reference point on the energy scale). Here $4\gamma$ is the energy bandwidth of the reservoir under consideration.

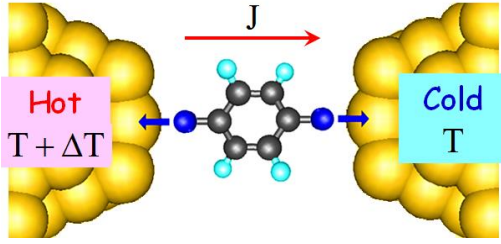

**Figure 2:** Schematic representation of molecular junction under investigation.

Figure 3 shows transmission function and its product with reflection function, because both of them are needed to perform calculations of transport characteristics and dynamical noise properties of nanoscale systems. We see that the most significant changes of mentioned functions are present at the vicinity of electrochemical potential (Fermi energy level) and therefore have direct impact on transport phenomena associated with electrons as heat carriers. In Fig.4, we plot linear thermal conductance and its nonlinear correction as functions of temperature. We noted a quadratic-to-linear transition in the temperature dependence of thermal conductance due to an increase of molecule-reservoir coupling in the range of intermediate temperatures. The nonlinear correction to thermal conductance starts at some finite value for $T \to 0$, increases to some maximum value, and then decreases linearly with temperature. From Fig.4, we see that the increase in molecule-reservoir coupling results in the shift of the maximum values of nonlinear corrections to thermal conductance (wide peaks) toward lower temperatures.

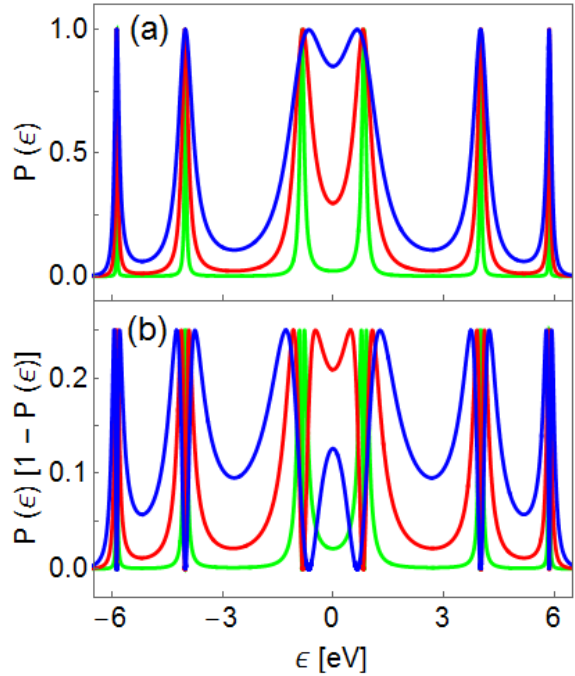

**Figure 3:** (a) Transmission function $P$ and (b) its product with reflection function $P[1-P]$ plotted against energy of electron $\varepsilon$ for different values of molecule-reservoir coupling strength (**green:** $\tau = 1$ eV, **red:** $\tau = 2$ eV, **blue:** $\tau = 3$ eV).

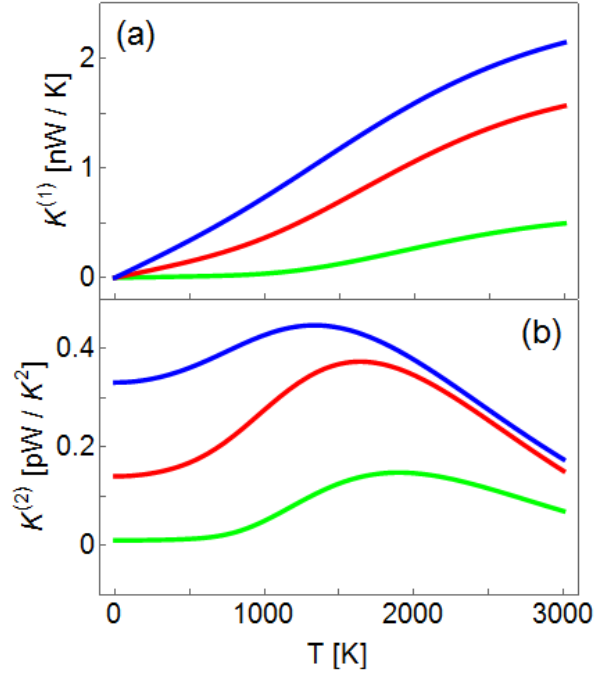

**Figure 4:** (a) Linear thermal conductance $K^{(1)}$ and (b) its nonlinear correction $K^{(2)}$ plotted against the temperature $T$ for different values of molecule-reservoir coupling strength (**green:** $\tau = 1$ eV, **red:** $\tau = 2$ eV, **blue:** $\tau = 3$ eV).

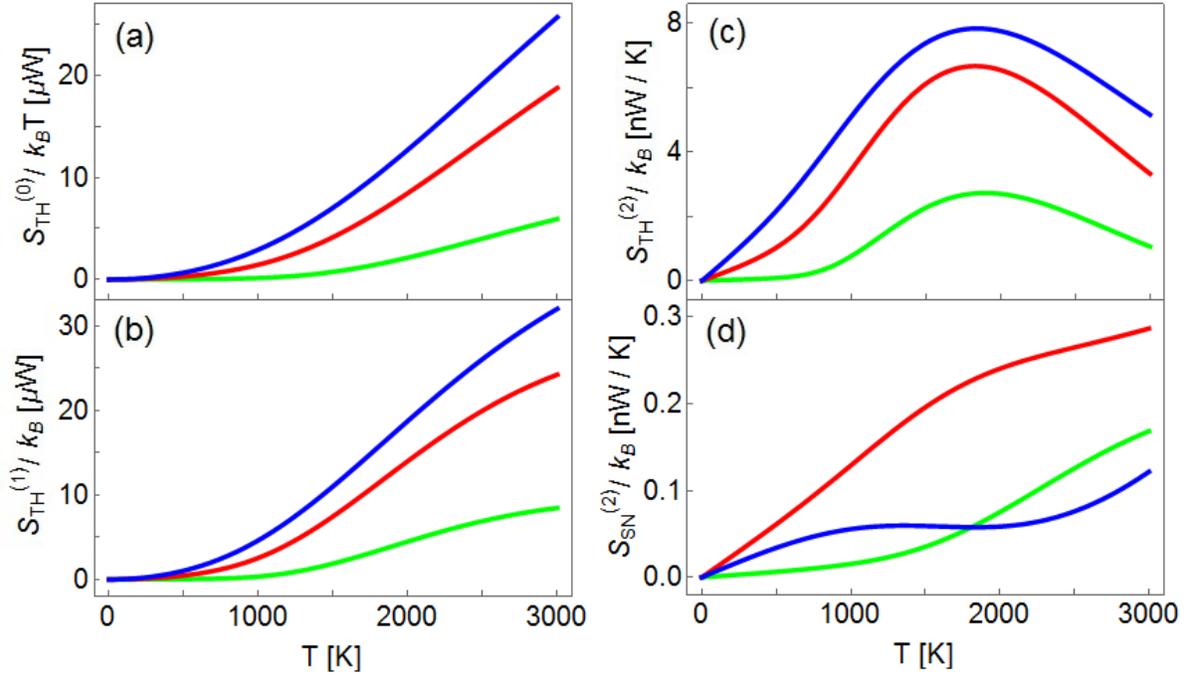

**Figure 5:** (a)-(c) Equilibrium thermal noise components $S_{TH}^{(0)}$, $S_{TH}^{(1)}$ and $S_{TH}^{(2)}$, and (d) the second-order non-equilibrium shot noise $S_{SN}^{(2)}$ plotted against the temperature $T$ for different values of molecule-reservoir coupling strength (**green:** $\tau = 1$ eV, **red:** $\tau = 2$ eV, **blue:** $\tau = 3$ eV).

Figure 5 shows dynamical noise properties of heat fluxes flowing through molecular junction under consideration. From this picture, we see that both zeroth-order and first-order thermal noises gradually increase with temperature. Further, the second-order thermal noise increases with temperature up to some maximum value, and then decreases linearly when temperature is increased above certain critical value. Interestingly, the temperature at which the second-order thermal noise reaches the maximum value only weakly depends on the molecule-reservoir coupling and for a set of parameters used in our model: $T_{max} \approx 1900$ K. The general trend is the following: the larger value of the molecule-reservoir coupling parameter, the greater values of all types of equilibrium thermal noises associated with nanoscale systems. Moreover, the second-order non-equilibrium shot noise strongly depends on the molecule-reservoir coupling and may be quite small even for large values of the molecule-reservoir coupling parameters, as clearly documented in Fig.5(d).

In statistical physics, the fluctuation-dissipation theorem (FDT) is a powerful tool applicable to both classical as well as quantum systems, being a direct consequence of the linear response theory. The FDT is limited to weak non-equilibrium conditions in which the linear response of the system (thermal conductance) onto relatively small perturbation (temperature difference) is related to the statistical behavior of that system (zeroth-order thermal noise) kept at equilibrium conditions (constant temperature). The quantitative expression for FDT is given via Eq.(30) and we checked numerically that it is valid for the molecular system under investigation for any temperature (results not shown). The importance of this particular formula stems from a simple fact that it is relatively easy to keep the system at constant temperature and conduct measurements at equilibrium thermodynamic conditions. Then we may use FDT to deduce the information about the linear reaction of the system onto applied temperature difference from a long-time observation of thermally activated fluctuations in that system (thermal noise).

However, it should also be noted that the fluctuation-dissipation theorem is violated in many nanoscale and mesoscopic systems, including glassy materials, proteins and low-dimensional quantum systems [31-36]. At strongly non-equilibrium conditions with high-intensity heat fluxes, the linear conductance and the zeroth-order thermal noise are insufficient to describe nonlinear behavior of molecular junctions of practical importance [37-54]. The question of how to extract the information about nonlinear corrections to transport and noise coefficients from our knowledge of their linear terms immediately arises. Certainly, it should be possible, since quantum transport in molecular systems and the associated thermal noise are both related to the same quantum-mechanical transmission function. Based on analysis performed in Section III, we can rewrite the expressions for higher-order (nonlinear) terms as linear functions of lower-order quantities and the average temperature of the analyzed system, namely

$$K^{(2)} = \frac{K^{(1)}}{2T} = \frac{S_{TH}^{(0)}}{8k_B T^3}, \qquad (49)$$

$$S_{TH}^{(1)} = 6k_B T K^{(1)} = \frac{3S_{TH}^{(0)}}{2T}, \qquad (50)$$

$$S_{TH}^{(2)} = 6k_B K^{(1)} = \frac{3S_{TH}^{(0)}}{2T^2}. \tag{51}$$

Substituting from Eq.(50) into Eq.(51), we obtain the noise-noise relation: $S_{TH}^{(1)} = TS_{TH}^{(2)}$. All those equations represent completely new set of noise-signal relations which interconnect thermal noise components to the appropriate transport coefficients. Those quantitative relations are of great importance in non-equilibrium quantum thermodynamics, because nonlinear corrections can be calculated by using equilibrium thermal noise and linear transport coefficient which are easily measurable quantities in comparison to any nonlinear terms. It should be noted that it is experimentally much easier to control the system at weak non-equilibrium situations than maintain strong non-equilibrium conditions for a long period of time (where the temperature difference plays the role of a driving force conjugate to the corresponding heat flux).

In Fig.6, we present numerical evidence that some of the noise-signal relations given by Eqs.(30)-(36) or by Eq.(49)-(51) are valid when temperature difference between heat reservoirs is significant, while the average temperature of the system under consideration is relatively small (in comparison to room temperature). It should be noted that transport via atoms and molecules is ballistic or quantum in nature and the dissipation of energy during scattering phenomena of heat carriers takes place inside the macroscopic reservoirs. All the noise-signal relations, discussed in this paper, couple statistical properties of a given system at equilibrium conditions (zeroth-order thermal noise) or during steady-state heat flow (higher-order noise terms) to relaxation processes described by exponential decay laws at non-equilibrium conditions (thermal conduction and its nonlinear corrections). In other words, there is the non-zero heat flux originated from thermally activated fluctuations inside particular reservoirs of molecular junctions which are kept at constant temperature. This heat flux is exactly the same as generated by relatively large temperature difference (perturbation) between those reservoirs. Here the characteristic relaxation times are identified with the reservoir-molecule escape rates, so they are closely related to the reservoir-molecule coupling parameters.

## VI. CONCLUDING REMARKS

Understanding transport of heat from the bottom-up perspective of quantum thermodynamics is of great scientific and technological importance. In particular, it is quite challenging to properly define and measure thermodynamic quantities for nanoscale and mesoscopic systems, ranging from simple molecules, quantum dots, single-electron devices, trapped-ion systems to colloids and biological settings. Thermodynamics becomes increasingly complicated as the system size decreases and fluctuations around average and steady-state quantities become significantly large. In those special cases, the probability of observing processes with entropy production opposite to that dictated by the second law of thermodynamics increases exponentially. Since the second law of thermodynamics is statistical in nature, there is non-zero probability that the entropy of isolated nanoscale system may spontaneously decrease for a short period of time. Although the questions of ergodicity and thermalization are beyond the scope of this paper, here we derived new noise-signal relations which are valid under the nearly-equilibrium steady-state conditions.

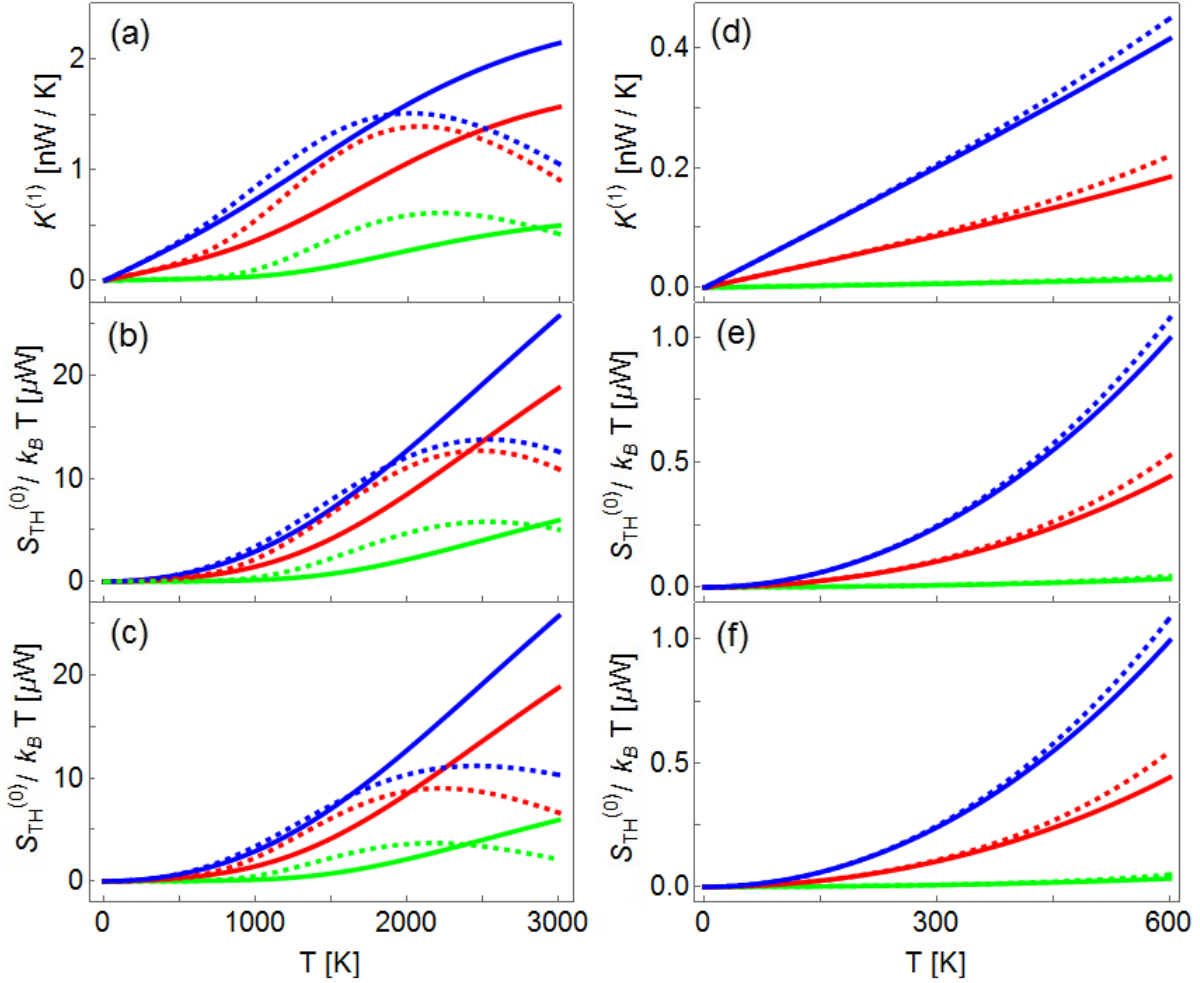

**Figure 6:** (a), (d) The exact solution for linear thermal conductance $K^{(1)}$ (solid line) and its approximated form $K^{(1)} = 2TK^{(2)}$ (dashed line) plotted against the average temperature of the molecular system under consideration. (b), (e) The exact solution for equilibrium thermal noise $S_{TH}^{(0)}$ (solid line) and its approximated form $S_{TH}^{(0)} = 8k_B T^3 K^{(2)}$ (dashed line) plotted against the average temperature of the molecular system under consideration. (c), (f) The exact solution for equilibrium thermal noise $S_{TH}^{(0)}$ (solid line) and its approximated form $S_{TH}^{(0)} = 2T^2 S_{TH}^{(2)} / 3$ (dashed line) plotted against the average temperature of the molecular system under consideration. All the results are presented for different values of molecule-reservoir coupling strength (**green:** $\tau = 1$ eV, **red:** $\tau = 2$ eV, **blue:** $\tau = 3$ eV).

Our original derivation is based on scattering (Landauer) approach to nanoscale heat conduction processes carried by electrons (fermionic heat carriers). Since the signal (heat flux) and the associated noise (dynamical noise power) are equally important in nanoscale systems, both quantities are treated on equal footing. Specifically, we proved that the ballistic Fourier's law and the fluctuation-dissipation theorem are applicable to nanoscale systems under any thermal conditions as long as temperature difference between heat reservoirs is relatively small in comparison to other energies in the analyzed system (where linear response theory is still valid).

Further, we derived a few asymptotic noise-signal relations with only limited applicability in non-equilibrium quantum thermodynamics. Such quantitative expressions allow one to calculate higher-order transport and noise terms on the basis of linear thermal conductance and equilibrium thermal noise. All those equations should be of great importance to scientists, who constantly deal with out-of-equilibrium open systems at molecular or mesoscopic scales [55].

It should be noted that practical realization of molecular junctions is always a compromise between the preservation of molecular electronic structure (resonant tunneling) and the formation of highly conductive connections (quasi-ballistic coupling). The simplicity of electronic structure of simple molecules (such as benzene) coupled to heat reservoirs may help in verification of existing models and validation of various approximations used to simplify calculations. In macroscopic systems, the more scattering we have in the system, the more noise we will measure by our apparatus. Following this logic, the ballistic heat conduction should be noiseless. However, noise in nanoscale systems is usually as important as signal itself, mainly due to quantum effects, such as: Pauli exclusion principle, interference of electronic waves, Coulomb blockade, etc. Here noise may be viewed as discrepancy from the actual result or time-dependent fluctuations which mask the measured signal (heat flux). Furthermore, the detailed analysis of noise provides useful information about statistical independence of heat carriers under consideration [56]. The shot noise spectrum contains the information about the quantum correlations between fermionic heat carriers (electrons), which lead to collective phenomenon of anti-bunching (see the Appendix C).

Recently, the experimental techniques to measure shot noise were used in order to study transport properties of molecular complexes [57-62]. The shot noise spectrum, as the second moment of the heat flux, provides additional information about nanoscale transport which is inaccessible via conventional measurements of thermal conductance. At nearly equilibrium conditions (low temperature and small bias voltage) and weak molecule-reservoir coupling regime, the measurements of shot noise allow one to estimate the whole spectrum of transmission function as well as the number of current-carrying conduction channels (number of modes). In this particular case, the thermal conductance of a coherent quantum system may be represented as a sum of independent conduction channels [58,59]. Combining shot noise and conductance measurements, it is also possible to study the effect of phonon activation on nanoscale transport in a view of the distribution of independent conduction channels, where the appearance of conductance enhancement may be a consequence of strong electron-phonon coupling or inter-channel scattering [60]. Importantly, the formalism presented in this paper is valid in the case of strong molecule-reservoir coupling regime, where all the interference and hybridization phenomena are included.

**APPENDIX A: Higher-Order Corrections**

In this appendix, we provide the expressions for temperature-dependent integral kernels that may be used to calculate the third and fourth order corrections with respect to temperature difference $\Delta T$. Those higher-order terms as related to electronic heat flux may be calculated by using Eq.(11) with the following functions

$$F_3(y) = \frac{k_B y^2}{16\pi\hbar T^2}\cosh^{-4}\left(\frac{y}{2}\right)\left[\left\{\frac{y^2}{6}+1\right\}\cosh(y)-y\sinh(y)-\frac{y^2}{3}+1\right], \tag{A1}$$

$$F_4(y) = \frac{k_B y^2}{16\pi\hbar T^3}\cosh^{-2}\left(\frac{y}{2}\right)\left[\left\{\frac{y^3}{12}+3y\right\}\tanh\left(\frac{y}{2}\right)-y^2-2\right.$$

$$\left.+\left\{\frac{3y^2}{2}-4y^3\tanh\left(\frac{y}{2}\right)\right\}\cosh^{-2}\left(\frac{y}{2}\right)\right], \tag{A2}$$

where variable $y=(\varepsilon-\mu)/k_B T$ as in Eq.(14). Further, the higher-order corrections to the equilibrium thermal noise associated with electronic heat flow may be calculated by using Eq.(25) with the following functions

$$\Phi_3(y) = \frac{k_B^2 y^3}{16\pi\hbar T}\cosh^{-5}\left(\frac{y}{2}\right)\left[\left\{1-\frac{11y^2}{6}\right\}\sinh\left(\frac{y}{2}\right)+\left\{\frac{y^2}{6}+1\right\}\sinh\left(\frac{3y}{2}\right)\right.$$

$$\left.+3y\cosh\left(\frac{y}{2}\right)-y\cosh\left(\frac{3y}{2}\right)\right], \tag{A3}$$

$$\Phi_4(y) = \frac{k_B^2 y^3}{16\pi\hbar T^2}\cosh^{-4}\left(\frac{y}{2}\right)\left[\left\{\frac{y^3}{6}+6y\right\}\cosh^2\left(\frac{y}{2}\right)+\frac{5y^3}{4}\cosh^{-2}\left(\frac{y}{2}\right)\right.$$

$$\left.-\{y^2+2\}\sinh(y)+6y^2\tanh\left(\frac{y}{2}\right)-\frac{5y^3}{4}-9y\right]. \tag{A4}$$

where variable $y=(\varepsilon-\mu)/k_B T$ as in Eq.(14). Finally, the higher-order corrections to the non-equilibrium shot noise associated with electronic heat flow may be calculated by using Eq.(38) with the following functions

$$\Psi_3(y) = \frac{k_B^2 y^4}{8\pi\hbar T}\cosh^{-4}\left(\frac{y}{2}\right)\left[\frac{y}{2}\tanh\left(\frac{y}{2}\right)-1\right], \tag{A5}$$

$$\Psi_4(y) = \frac{3k_B^2 y^4}{32\pi\hbar T^2}\cosh^{-6}\left(\frac{y}{2}\right)\left[\left\{\frac{7y^2}{36}+1\right\}\cosh(y)-y\sinh(y)-\frac{11y^2}{36}+1\right]. \tag{A6}$$

where variable $y=(\varepsilon-\mu)/k_B T$ as in Eq.(14).

**APPENDIX B: Quantum of Thermal Conductance**

In this appendix, we present a simple justification behind the concept of thermal conductance quantization by combining the equipartition theorem from thermodynamics and Heisenberg uncertainty principle from quantum theory. According to the equipartition theorem, the average energy associated with each degree of freedom of the system under investigation is $E = k_B T/2$, hence the change of the energy during the process of heat transfer is simply

$$\Delta E = \frac{k_B}{2} \Delta T. \tag{B1}$$

The heat flux is defined as the amount of thermal energy transferred from one heat reservoir to the other via quantum system within certain time interval $\tau$, therefore $J = k_B T/2\tau$. From that relation, we can extract the following expression for the characteristic time

$$\tau = \frac{k_B T}{2J}. \tag{B2}$$

By virtue of Heisenberg uncertainty principle for complementary variables of energy and time, we can write down the following inequality

$$\Delta E \cdot \tau \geq \frac{\hbar}{2}. \tag{B3}$$

Substituting from Eq.(B1) and Eq.(B2) into Heisenberg inequality from Eq.(B3), we obtain

$$\frac{k_B^2 T}{4J} \Delta T \geq \frac{\hbar}{2}. \tag{B4}$$

By taking into consideration the formal definition of the linear thermal conductance as heat flux divided by temperature difference $K \equiv J/\Delta T$, Eq.(B4) may be rewritten in the form

$$K \leq \frac{k_B^2 T}{2\hbar}. \tag{B5}$$

From the above inequality, we see that the maximum value for the linear thermal conductance during the process of heat transport as carried by electrons via single energy level (or carried by acoustic phonons associated with individual vibrational mode) is very close to the quantum of thermal conductance $K^{(1)} = \pi k_B^2 T/6\hbar$ derived from the Landauer formula in section II (the difference between two results is less than 5%).

**APPENDIX C: Analysis of Fano Factor**

In this appendix, we perform analysis of Fano factor for purely ballistic transport regime. Fist of all, it should be noted that electrons are intrinsically noisy due to thermal phenomena and quantum effects. Thermal noise is a direct consequence of random motion of heat carriers because of non-zero temperature of the sample, while quantum (shot) noise is closely related to Heisenberg uncertainty principle and Pauli exclusion principle for fermionic heat carriers (like electrons). In Section III, we showed that the stochastic nature of heat reservoir generates thermal and quantum fluctuations of heat fluxes which are not arbitrary, but these quantities are interrelated via the special type equations, known as noise-signal relations. In particular, shot noise determines dynamic fluctuations of heat fluxes due to granularity of electrons (corpuscular property). Furthermore, the detailed analysis of noise provides information, which is unavailable via measurements of heat fluxes, about statistical independence of heat carriers under consideration. The steady-state heat flux is a measure of average energy transferred by individual carriers, while noise spectrum offers information about statistics of electrons and their quantum correlations leading to collective phenomenon of anti-bunching.

The Fano factor is defined as a dimensionless noise-to-signal ratio. In the case of linear response theory, the Fano factor may be defined as follows

$$F \equiv \frac{S_{TH}^{(0)}}{4k_B T^2 K^{(1)}} = 1. \tag{C1}$$

The statistical independence of electrons in the linear ballistic transport regime is reflected in the Poissonian value for Fano factor $F = 1$. Eq.(C1) is actually a simple demonstration of the fluctuation-dissipation theorem (FDT). However, in the case of nonlinear ballistic transport regime, the heat flux is expressed via Eq.(15) and the Fano factor with up to the second order noise terms may be defined as

$$F \equiv \frac{S_{TH}^{(0)} + S_{TH}^{(1)} \Delta T + S_{TH}^{(2)} (\Delta T)^2 + S_{SN}^{(2)} (\Delta T)^2}{4k_B T^2 (K^{(1)} + K^{(2)} \Delta T)} = \sum_{m=0}^{3} F^{(m)}. \tag{C2}$$

In the case of collisionless transport of electrons flowing via single energy level we have $P(\varepsilon) = 1$ and the additive terms of the Fano factor from Eq.(C2) may be calculated analytically

$$F^{(0)} = \frac{S_{TH}^{(0)}}{4k_B T^2 (K^{(1)} + K^{(2)} \Delta T)} = \frac{2}{2+x}, \tag{C3}$$

$$F^{(1)} = \frac{S_{TH}^{(1)} \Delta T}{4k_B T^2 (K^{(1)} + K^{(2)} \Delta T)} = \frac{3x}{2+x}, \tag{C4}$$

$$F^{(2)} = \frac{S_{TH}^{(2)} (\Delta T)^2}{4k_B T^2 (K^{(1)} + K^{(2)} \Delta T)} = \frac{3x^2}{2+x}. \tag{C5}$$

Here the dimensionless variable $x = \Delta T / T$ and in this particular case we obtain $S_{SN}^{(2)} = 0$ (and consequently $F^{(3)} = 0$). If we add together Eqs.(C3)-(C5), we obtain the temperature-dependent expression for the minimum value of the Fano factor

$$F_{min} = \sum_{m=0}^{2} F^{(m)} = \frac{2 + 3x + 3x^2}{2+x}. \tag{C6}$$

Interestingly, thermal noise is statistics-independent quantity and turned out to be exactly the same for both fermionic and bosonic heat carriers (like electrons and phonons) [63]. However, the non-equilibrium shot noise generated by electrons differs from this generated by acoustic phonons not only by its sign, but also by its magnitude [64]. In general, the electronic shot noise is non-negative and usually increases the value of the Fano factor, while the phononic shot noise is non-positive and usually reduces the value of the Fano factor.

As mentioned in Section III, the non-equilibrium shot noise is a non-negative quantity that reaches the maximum value for transmission function $P(\varepsilon) = 1/2$. In this specific situation, both heat flux as well as thermal noise are reduced by half, but their ratio still remains the same. As a consequence, the thermal noise components of the Fano factor will not change, while the shot noise component $F^{(3)}$ may be calculated as

$$F^{(3)} = \frac{S^{(2)}_{SN,max}(\Delta T)^2}{2k_B T^2 (K^{(1)} + K^{(2)} \Delta T)} = \frac{x^2/2}{2+x}. \tag{C7}$$

If we add together Eqs.(C3)-(C5) and Eq.(C7), we obtain the temperature-dependent expression for the maximum value of the Fano factor

$$F_{max} = \sum_{m=0}^{3} F^{(m)} = \frac{2 + 3x + 7x^2/2}{2+x}. \tag{C8}$$

According to the spin-statistics theorem, and as a consequence of quantum correlation effects between electrons, the electronic wave function describing ensemble of fermionic particles must be antisymmetric with respect to exchange of two fermions. Moreover, the equilibrium distribution function of fermionic carriers takes the form of Fermi-Dirac statistics. However, at non-equilibrium conditions, electrons tend to avoid each other due to their anti-bunching property. Such tendency ultimately results in positive shot noise Fano factor $F^{(3)} > 0$ which increases super-Poissonian noise power spectral density above the level established by thermal fluctuations. It should be noted that the shot noise becomes dominant when the finite number of electrons is sufficiently small, so that uncertainties associated with Poissonian distribution become significant.

As shown in Fig.7, the Fano factor reaches its Poissonian limit $F = 1$ at equilibrium thermal conditions at which $\Delta T \to 0$. In this particular case, there is no correlations between electrons which are moving freely through the sample. However, at non-equilibrium conditions, when $\Delta T > 0$, we obtain positive value of shot noise Fano factor $F^{(3)} > 0$ within the super-Poissonian regime of the total noise power $F > 1$. Here the enhancement of thermal super-Poissonian noise is due to the so-called anti-bunching effect of electrons. This anti-bunching property of fermionic heat carriers is a direct consequence of Pauli exclusion principle which is a natural tendency of fermions to avoid the same quantum state (and consequently each other). Note that the non-equilibrium shot noise constitutes a relatively small contribution to the total noise power which is entirely dominated by thermal fluctuations even in the case of large temperature difference between heat reservoirs.

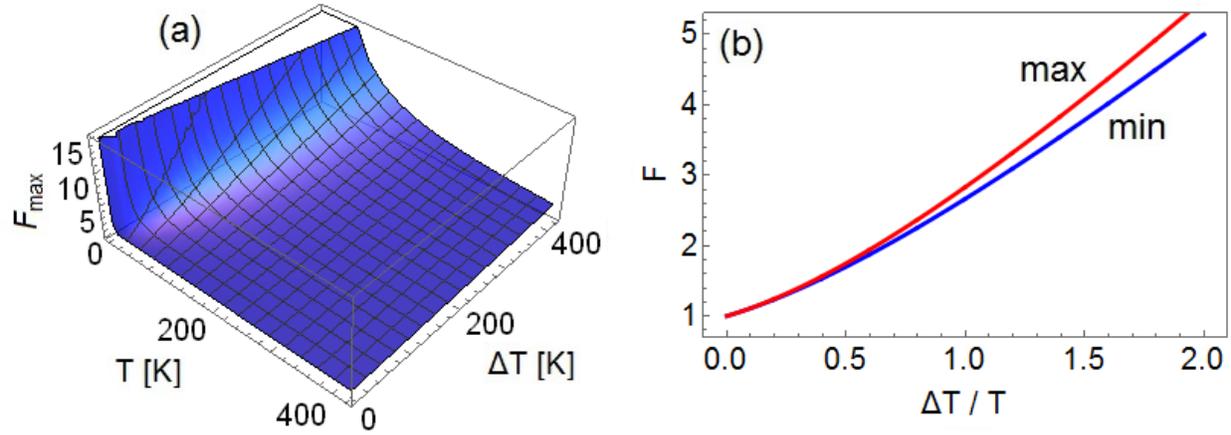

**Figure 7:** (a) 3D plot of maximum value of dimensionless Fano factor for ballistic transport of electrons as a function of absolute temperature $T$ and temperature difference $\Delta T$; (b) 2D plot of Fano factor F as a function of ratio $\Delta T/T$ (indicated both its minimum and maximum values).

**ACKNOWLEDGMENTS**

This work was supported by Pace University Start-up Grant. The authors are grateful to Angelika Walczak and Demos Athanasopoulos for many valuable discussions.